# Software defined radio layer for IR-UWB systems in Wireless Sensor Network Context

Aubin Lecointre
LAAS-CNRS, University of Toulouse / 7, Av du Colonel Roche, 31077 Toulouse cedex 4/ France
alecoint@laas.fr

Daniela Dragomirecu
LAAS-CNRS, University of Toulouse / 7, Av du Colonel Roche, 31077 Toulouse cedex 4/ France
daniela@lass.fr

Robert Plana
LAAS-CNRS, University of Toulouse / 7, Av du Colonel Roche, 31077 Toulouse cedex 4/ France
plana@lass.fr

*Abstract* – **This paper addresses the radio interface problematic for MANET (Mobile Ad-hoc NETwork) applications. Here we propose to study the radio reconfigurability in order to provide a unique physical layer which is able to deal with all MANET applications. For implementing this reconfigurable physical layer, we propose to use Impulse Radio Ultra WideBand (IR-UWB). This paper presents also our two level design approach for obtaining our reconfigurable IR-UWB receiver on FPGA (Field Programmable Gate Array).**

## I. INTRODUCTION

In this paper we focus on MANets (Mobile Ad-hoc Netwotks) in the Wireless Sensors Networks (WSN) context. This context implies some particulars constraints on the radio interface, such as the size, the cost, the simplicity and the energy needs of the proposed architecture. IR-UWB (Impulse Radio – Ultra Wide Band) [1] is a very promising technology for this kind of applications, i.e. short range wireless data applications. IR-UWB has some advantages: 7,5 GHz of free spectrum which could permit to reach high data rate, extremely low transmission energy, extremely difficult to intercept, multi-path immunity, low cost (mostly digital architecture), "Moore's Law Radio" (performances, size, data rate, cost follow Moore's Law), simple CMOS transmitter at very low power [2]. We will expose the IR-UWB (Impulse Radio – Ultra WideBand) modulation which seems to be viable in this context: Time Hopping used with OOK (On Off Keying), BPAM (Binary Pulse Amplitude Modulation), or PPM (Pulse Position Modulation). We will propose here a reliable and software defined radio layer for ad hoc smart dust networks. We will take into account the problem of radio interface for smart dust systems on chip for ad-hoc networks. Results introduced here will be implemented on a FPGA (Field Programmable Gate Array) circuits in order to answer to the WSN-MANet problematic.Thus in this paper, we will expose how we could design FPGA circuit for obtaining a data rate and TH-code reconfigurable receiver; and we will characterize this solution according to reliability (BER versus SNR criteria), data rate and WSN constraints.

This paper is laid out as follow: section II introduces our design approach, section III presents the high level modelling and validation of our distinct receiver architectures using Matlab, section IV shows FPGA implementation of our IR-UWB solutions and section V introduce the radio reconfigurability concept and its implementation for an IR-UWB receiver on FPGA.

## II. DESIGN APPROACH

In order to obtain a reconfigurable receiver, we have followed a two levels (system and hardware) development way as described in figure 1.

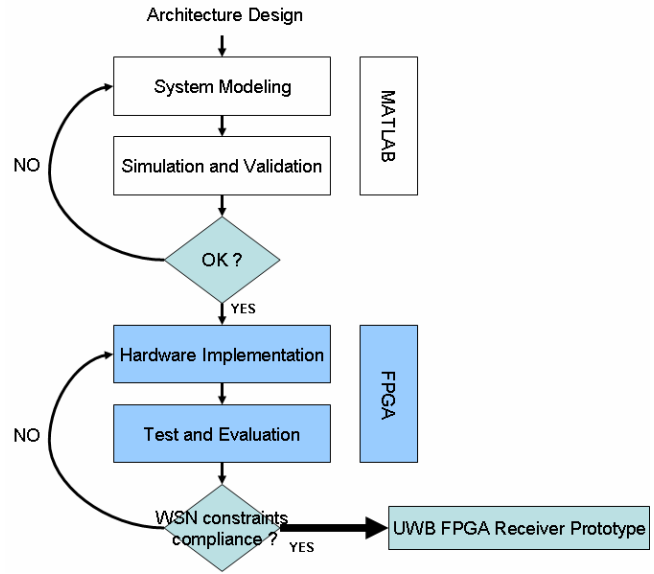

Figure 1. The two level development approach.

The first level consists in designing and modeling the receiver architecture on Matlab. This system level modeling permits us to validate the viability of the proposed architectures in the MANET context. Furthermore, our Matlab platform is also used for performance characterization according different criteria, such as BER versus SNR, or MANET/WSN constraints. If the performances are good enough, we can start the hardware level implementation on FPGA. The second level permit to achieve our goal: designing an IR-WUB reconfigurable receiver on FPGA. We can see in Figure 1 two feed back link which embody the possibility of re-adjust the architecture if the performances don't satisfy our WSN constraints.

III. HIGH LEVEL MODELING OF UWB TRANSCEIVERS

We have developed with Matlab a full-parametric IR-UWB communication link model. This one includes emitter, receiver and channel modelling. Thanks to it, we have validated our receiver architecture for TH-PPM, TH-OOK, and TH-BPAM [1]. We use for TH-PPM a double correlation with template waveform coherent receiver [1] as described in Figure 2. This coherent receiver requires synchronization for performing correctly the correlation. This synchronization is carried out by a matched filter [3]. Nevertheless, this need implies an increase of the complexity of the architecture.

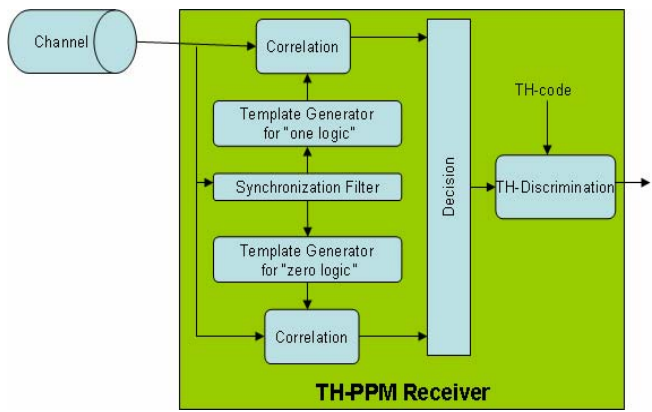

Figure 2.  TH-PPM coherent receiver.

TH-BPAM re-uses the same receiver principle except that it's a simple correlation. As a result this receiver is simpler than the TH-PPM receiver. The simplest solution is the TH-OOK receiver [4]. It consists in a non coherent receiver based on energy detection (Figure 3).

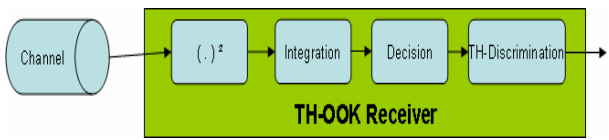

Figure 3.  TH-OOK non coherent receiver architecture.

Once these three architectures implemented, we have compared them according to the BER versus SNR criteria, and the WSN constraints. Table I summarizes the receiver classification. The BER-classification is obtained thanks to our Matlab platform. We have characterized theirs performances in the IEEE 802.15.4a channel (figure 4) [5]. For the WSN constraints we have obtained the classification by analyzing our three architectures.

Figure 4 includes UWB pulse modulations techniques performances. We could see that IR-UWB systems offer viable performances for our context of applications. Moreover figure 4 proves the table 1 results i.e. TH-PPM is better than TH-PAM, which is better than TH-OOK.

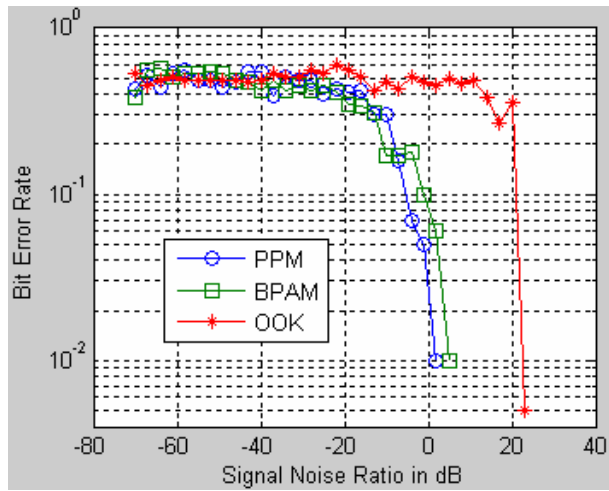

Figure 4.  IR-UWB modulation techniques according to evaluation BER/SNR criteria, on IEEE 802.15.4a UWB channel.

| Classification | WSN Constraints | | | | Performances |
|---|---|---|---|---|---|
| IR-UWB for WSN | Power | Cost | Simplicity | Size | BER vs SNR |
| TH-PPM | 3 | 3 | 2 | 3 | 1 |
| TH-BPAM | 2 | 2 | 2 | 2 | 2 |
| TH-OOK | 1 | 1 | 1 | 1 | 3 |

TABLE I.  COMPARATIVE ANALYSIS OF IR-UWB ARCHITECTURES

Using high level modelling with Matlab, we have demonstrated the viability, for the MANet WSN context, of three IR-UWB receiver architectures; we have classified them, and validate their operation. Besides, table 1 proves the existence of a trade-off between the respect of the WSN constraint and the BER performance. Now we are able to implement them at hardware level, on a FPGA.

IV. FPGA DESIGN OF AN UWB RECEIVER

We have designed distinct Xilinx FPGA receivers based on IR-UWB techniques with good hardware results. Nevertheless, in this paper, we will focus on data rate and TH-code reconfigurable IR-UWB receiver. The radio reconfigurability mechanism is also applicable for others receivers (TH-OOK and TH-BPAM).

Figure 5 presents the co-simulation and co-performance evaluation platform for IR-UWB FPGA receiver. Matlab simulate the MAC layer, the emitter, channel and ADC and the FPGA implements the digital receiver part. This platform allows designing, simulating, implementing and evaluating the BER and WSN performances of the FPGA receiver.

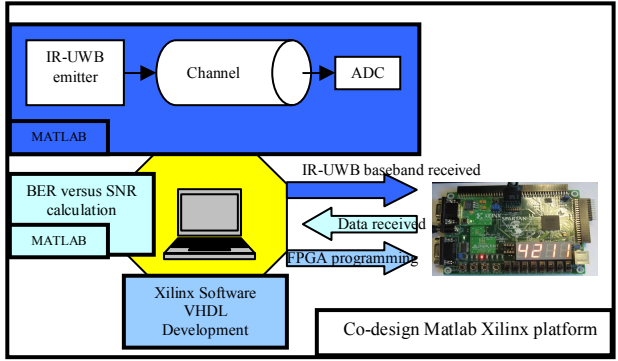

Figure 5. Co-simulation and co-performance platform for IR-UWB FPGA receiver.

We have implemented eight versions of our receivers. They are resumed in the table II. There are five parameters for implementations: sample size, number of channel, presence of localization mechanism and reconfigurability, and bloc position. In addition to allow us to establish a relative comparison of our implementations, the hardware prototyping will permit us to determine the impact of each parameter.

| VERSION | Receiver Properties ||||| 
|---|---|---|---|---|---|
| | Receiver Principle | Sample Size | Channel Number | Distance | Radio Reconf |
| TH-OOK-v1 | Energy Detection | 64 bits | Mono Channel | NO | NO |
| TH-OOK-v2 | Energy Detection | 32 bits | Mono Channel | NO | NO |
| TH-BPAM-v1 | Simple Correlation | 32 bits | Mono Channel | NO | NO |
| TH-BPAM-v2 | Simple Correlation | 32 bits | Mono Channel | NO | NO |
| TH-PPM-v1 | Double Correlation | 32 bits | Mono Channel | NO | NO |
| TH-PPM-v2 | Double Correlation | 32 bits | Mono Channel | YES | NO |
| TH-PPM-v3 | Double Correlation | 64 bits | Mono Channel | NO | YES |
| TH-PPM-v4 | Double Correlation | 64 bits | Double Channel | NO | YES |

TABLE II. IR-UWB RECEIVER IMPLEMENTATIONS ON FPGA

| VERSION | WSN Constraints || Performances |
|---|---|---|---|
| | Size (in gate) | Max. Frequency | BER/SNR |
| TH-OOK-v1 | 86 234 + 864 | 89,506 MHz | 8th |
| TH-OOK-v2 | 68 666 +864 | 111,693 MHz | 7th |
| TH-BPAM-v1 | 68 663 + 864 | 111,693 MHz | 5th |
| TH-BPAM-v2 | 68 716 +864 | 111,693 MHz | 5th |
| TH-PPM-v1 | 69 674 + 864 | 110,9 MHz | 1st |
| TH-PPM-v2 | 101 565 + 864 | 66,089 MHz | 1st |
| TH-PPM-v3 | 107 753 + 864 | 72,844 MHz | 3rd |
| TH-PPM-v4 | 125 441 + 1056 | 72,844 MHz | 3rd |

TABLE III. RECEIVERS' COMPARISON.

Table III summarizes the performances of our FPGA receiver according the WSN constraints and the BER versus SNR criteria. This table gives us an idea of the relative performances of our architectures, and not absolute performances, since we are limited by the FPGA, here a Xilinx Spartan 3.

We notice that they offer different performances. Some versions are most convenient for high data rate (maximum frequency criteria) applications and small size requirements, while others embedded complex functionality such as ranging and reconfigurability which explains theirs lower performances (regarding size and frequency constraints).

## V. FPGA RECONFIGURABIBLITY IMPLEMENTATION

We have implemented properties reconfigurable receiver. Each receiver is able to change its properties in order to satisfy application needs, as the opposite of the architecture reconfigurable concept which proposes to change the architecture instead of properties. Figure 6 illustrates the two visions of reconfigurability.

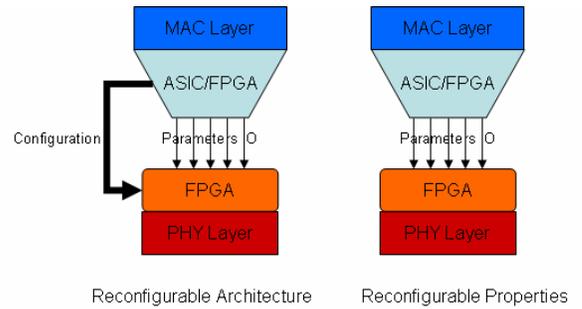

Figure 6. Illustration of the two vision of reconfigurability.

This is a dynamic reconfigurable receiver because it is able to change its properties, while the system is in used. The parameter change for the reconfigurability is decided by MAC layer and send to FPGA via Matlab simulation platform. For implementing the software defined radio concept (reconfigurability), we parameterize receiver properties, at VHDL level, in data rate and TH-code. The data rate is function of time hopping parameters such as frame duration (Tf), slot duration (Tc), and number of slot per frame (Nc):

$$D_{total}(bits/s) = \frac{N_c}{T_f} = \frac{N_c}{Nc \times T_c} = \frac{1}{T_c}$$

We have to implement Tc, and Nc as VHDL entries entity in order to obtain data rate reconfigurability. Thus for changing the data rate, we have to change the value of the slot duration (Tc).

The same principle is set up for TH-code reconfigurability. As TH-codes are digital values, thus they could be saved in memories. TH-code reconfigurability could be defined as a change of these values at the correct time. In order to implement it, we have designed a TH-code management bloc, added in the IR-UWB receiver, as described in Figure 7.

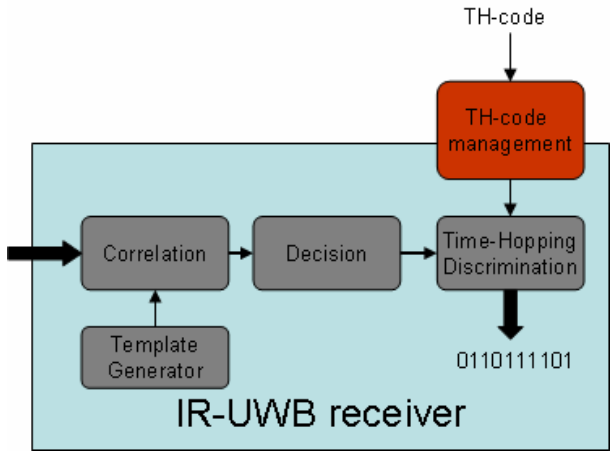

Figure 7.  Implementation of TH-code reconfiguration in TH-BPAM receiver.

Figure 8 illustrates our reconfigurable parameters implementations as input of the PHY layer and as output of the MAC layer. We have added a reconfiguration signal in order to activate the reconfiguration when correct values are placed on PHY layer inputs.

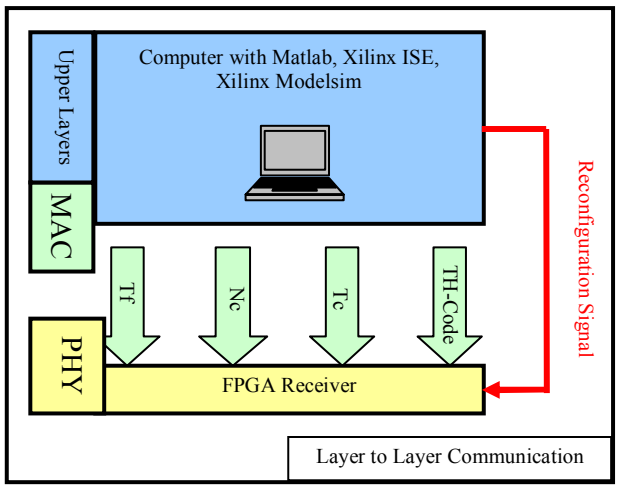

Figure 8.  PHY and MAC layer interface : implementation of reconfigurable parameters

This technique has some limitations due to the limit size of the VHDL entity entries. Entry size will establish the achievable data rate but we will also increase the size and the consumption of the circuit. These two criteria (data rate on one hand and size and consumption on the other) are opposed to each other and a compromised have to be found depending on the application.

Table III presents the comparative performance of our IR-UWB FPGA receivers' implementations with respect to the important criteria in WSN: the size, the data rate (frequency) and the BER. Reconfigurable and non-reconfigurable FPGA receiver architectures are compared in this table.

Figure 9 permits to visualize the RTL schematic view of the TH-PPM-v4 receiver. This receiver is a static reconfigurable one. Thus Figure 9 allows us to describe the presence of the data rate reconfigurable parameters: chip duration (Tc) and the number of chip per frame (Nc). On Figure 9, we can see that Tc and Nc are implemented as PHY layer inputs. Moreover, we can also see the TH-code management bloc which is responsible for the TH-code reconfiguration.

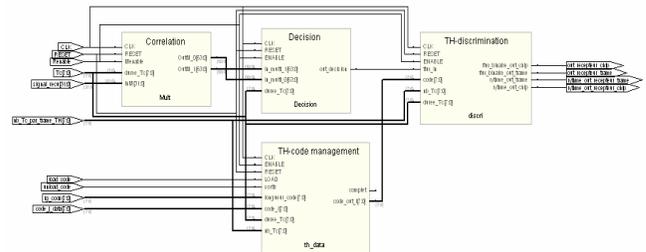

Figure 9.  RTL schematic view of a reconfigurable IR-UWB TH-PPM receiver on FPGA

## VI. CONCLUSION

High level modelling, circuit level implementation and performance evaluation of IR-UWB receivers are present in this paper. The co-design is used to propose a data rate and TH-code reconfigurable receiver and to establish its viability in the Wireless Sensor Network context.

REFERENCES

[1] I.Opperman, Jari Iinatti, Matti Hčamčalčainen, « UWB theory and applications », *Wiley* 2004.
[2] D. Morche, C. Dehos, T. Hameau, D. Larchartre, M. Pelissier, D. Helal, L. Smaini, « Vue d'ensemble des architecture RF pour l'UWB », LETI, UWB Summer School , Valence, France, oct. 2006 à l'ESISAR.
[3] MG. Di Benedetto, L. De Nardis, M. Junk, G. Giancola, « (UWB)[2]: Uncoordinated, Wireless, Baseborn Medium Access for UWB Communication Networks", *Mobile Networks and Applications*, vol. 10, no. 5, October 2005.
[4] LM Aubert, Ph.D. disertation: "Mise en place d'une couche physique pour les futurs systèmes de radiocommunications hauts débits UWB », INSA Rennes, France,  2005.
[5] A. Molisch, et al., « IEEE 802.15.4a channel model – final report », IEEE 802.15.4a.